\documentclass[preprint]{aastex}
\usepackage{epsf}
\usepackage{emulateapj5}
\usepackage{onecolfloat}
\usepackage{apjfonts}
\usepackage{amsmath}

\def \xoff {\ifmmode x_{\rm off} \else $x_{\rm off}$ \fi}
\def \rhorms {\ifmmode \rho_{\rm rms} \else $\rho_{\rm rms}$ \fi}

\def \apjs  {ApJS}

\def \mnras {MNRAS}
\def \etal {et~al.~}
\def \chisq  {\ifmmode  \chi^2   \else  $\chi^2$  \fi}
\def \chisqr {\ifmmode \chi^2_{\rm r} \else $\chi^2_{\rm r}$ \fi}
\def \spose#1{\hbox  to 0pt{#1\hss}}
\def \lta{\mathrel{\spose{\lower 3pt\hbox{$\sim$}}\raise  2.0pt\hbox{$<$}}}
\def \gta{\mathrel{\spose{\lower  3pt\hbox{$\sim$}}\raise 2.0pt\hbox{$>$}}}

\def \ha  {\ifmmode H\alpha \else H$\alpha $ \fi}

\def \kms {\ifmmode  \,\rm km\,s^{-1} \else $\,\rm km\,s^{-1}  $ \fi }
\def \kpc {\ifmmode  {\rm kpc}  \else ${\rm  kpc}$ \fi  }
\def \Msun {\ifmmode M_{\odot} \else $M_{\odot}$ \fi}
\def \hMsun {\ifmmode h^{-1}\,\rm M_{\odot} \else $h^{-1}\,\rm M_{\odot}$ \fi}
\def \hhMsun {\ifmmode h^{-2}\,\rm M_{\odot}\else $h^{-2}\,\rm M_{\odot}$ \fi}
\def \Lsun {\ifmmode L_{\odot} \else $L_{\odot}$ \fi}
\def \hhLsun {\ifmmode h^{-2}\,\rm L_{\odot} \else $h^{-2}\,\rm L_{\odot}$ \fi}

\def \LCDM {\ifmmode \Lambda{\rm CDM} \else $\Lambda{\rm CDM}$ \fi}
\def \sig8 {\ifmmode \sigma_8 \else $\sigma_8$ \fi}
\def \OmegaM {\ifmmode \Omega_{\rm M} \else $\Omega_{\rm M}$ \fi}
\def \OmegaL {\ifmmode \Omega_{\rm \Lambda} \else $\Omega_{\rm \Lambda}$\fi}
\def \Deltavir {\ifmmode \Delta_{\rm vir} \else $\Delta_{\rm vir}$ \fi}

\def \rs {\ifmmode r_{\rm s} \else $r_{\rm s}$ \fi}
\def \rrm2 {\ifmmode r_{-2} \else $r_{-2}$ \fi}
\def \ccm2 {\ifmmode c_{-2} \else$c_{-2}$ \fi}
\def \cvir {\ifmmode c_{\rm vir} \else $c_{\rm vir}$ \fi}
\def \cbar {\ifmmode \overline{c} \else $\overline{c}$ \fi}

\def \R200 {\ifmmode R_{200} \else $R_{200}$ \fi}
\def \Rvir {\ifmmode R_{\rm vir} \else $R_{\rm vir}$ \fi}

\def \v200 {\ifmmode V_{200} \else $V_{200}$ \fi}
\def \Vvir {\ifmmode V_{\rm  vir} \else  $V_{\rm vir}$  \fi}
\def  \Vhalo  {\ifmmode V_{\rm halo} \else $V_{\rm halo}$ \fi}

\def \M200 {\ifmmode M_{200} \else $M_{200}$ \fi}
\def \Mvir {\ifmmode M_{\rm  vir} \else $M_{\rm  vir}$ \fi}
\def \Mshell  {\ifmmode M_{\rm shell} \else $M_{\rm shell}$ \fi}

\def \Nvir {\ifmmode N_{\rm  vir} \else $N_{\rm  vir}$ \fi}

\def \Jvir {\ifmmode J_{\rm vir} \else $J_{\rm vir}$ \fi}
\def \Jshell {\ifmmode J_{\rm shell} \else $J_{\rm shell}$ \fi}

\def \Evir {\ifmmode E_{\rm vir} \else $E_{\rm vir}$ \fi}

\def \lam {\ifmmode \lambda  \else $\lambda$ \fi}
\def \lamp {\ifmmode \lambda^{\prime} \else $\lambda^{\prime}$  \fi}
\def \lampc {\ifmmode \lambda^{\prime}_{\rm c} \else
  $\lambda^{\prime}_{\rm c}$  \fi}
\def \lambar {\ifmmode \bar{\lambda}  \else  $\bar{\lambda}$  \fi}
\def  \lampbar  {\ifmmode \bar{\lambda^{\prime}} \else
  $\bar{\lambda^{\prime}}$\fi}
\def \siglam {\ifmmode \sigma_{\lambda} \else $\sigma_{\lambda}$ \fi}
\def \siglamp {\ifmmode                \sigma_{\lambda^{\prime}} \else
$\sigma_{\lambda^{\prime}}$\fi}

\def\LCDM{$\Lambda$CDM }

\def\c200{$c_{200}$}

\begin{document}
\submitted{The Astrophysical Journal, submitted}
\vspace{1mm}
\slugcomment{{\em The Astrophysical Journal, submitted}}

\shortauthors{Zhao et al.} \twocolumn[ \lefthead{Structure Formation by Fifth Force}

\righthead{Zhao et al.}

\title{Structure Formation by Fifth Force: Power Spectrum from N-Body Simulations}

\author{Hongsheng Zhao$^{1,2}$, Andrea V. Macci\`o$^{3}$, Baojiu Li$^{4,5}$, Henk Hoekstra$^{2}$, Martin Feix$^{1}$}
\affil{$^1$SUPA, School of Physics and Astronomy, University of St Andrews, KY16 9SS, UK \\
$^2$ Leiden University, Leiden Observatory, Niels-Borhweg 2, 2333CA, Leiden, the Netherlands \\
$^3$ Max-Planck-Institut f\"ur Astronomie, K\"onigstuhl 17, 69117
  Heidelberg, Germany \\
$^4$ DAMPT, Centre for Mathematical Sciences, University of Cambridge, CB3 0WA, UK\\
$^5$ Kavli Institute for Cosmology Cambridge, Madingley Road, Cambridge CB3 0HA, UK}
\begin{abstract}

We lay out the framework to numerically study nonlinear structure formation in the context of scalar-field-coupled cold dark matter models ($\varphi$CDM models) where the scalar field $\varphi$ serves as dynamical dark energy.
Adopting parameters for the scalar field which leave negligible effects on the CMB spectrum,
we generate the initial conditions for our $N$-body simulations. The simulations
follow the spatial distributions of dark matter and the scalar field, solving their equations of motion using a
multilevel adaptive grid technique. We show that the spatial
configuration of the scalar field depends sensitively on the local density field.
The $\varphi$CDM model differs from standard $\Lambda$CDM at small scales with observable modifications of, e.g., 
the mass function of halos as well as the matter power spectrum.  Nevertheless, the predictions of both models for the Hubble expansion
and the CMB spectrum are virtually indistinguishable. Hence, galaxy cluster counts and weak lensing observations,  
which probe structure formation at small scales, are needed to falsify this class of models.

\end{abstract}

\keywords{cosmology: theory --- methods: $N$-body simulations}
]

\section{Introduction}
The origin and nature of dark energy (Copeland~et~al.~2006) is one of
the most difficult challenges facing physicists and cosmologists at
the present time. Among all the proposed models to tackle this
problem, the introduction of a scalar field is perhaps the most
popular.  The scalar field, denoted by $\varphi$, should have no
coupling to normal matter to be consistent with stringent constraints
from experiments (Will 2006, and references therein), but could couple to the dark matter, therefore
producing a fifth force between dark matter particles. This idea has
gained a lot of interest in recent years because dark matter physics
are unknown, and such a coupling could alleviate the coincidence
problem of dark energy (e.g., Amendola 2000; Chiba 2001; Chimento et al. 2003). Furthermore, it is commonly predicted by low
energy effective theories derived from a more fundamental theory. A specific and
interesting possibility is the chameleon mechanism (Khoury \& Weltman
2004; Mota \& Shaw 2006), by virtue of which the scalar field acquires
a large mass in high density regions and thus the fifth force becomes
undetectable on short ranges, thus also evading constraints from
the large-scale cosmic microwave background (CMB). Indeed, at the linear
perturbation level, there have been a lot of studies about the coupled
scalar field and $f(R)$ gravity models (e.~g., Li \& Barrow 2007; Hu
\& Sawicki 2007).

Nevertheless, little is known about these models on nonlinear
scales. It is well known that the matter distribution at late times, i.e. $z\lesssim 2$ for cluster scales,
evolves in a nonlinear way, making the behavior of the scalar field more
complex and the linear analysis insufficient to produce accurate
results that can be confronted with observations. For the latter purpose, the
best way forward is to perform full $N$-body simulations (Bertschinger
1998) to evolve the individual particles step by step.

$N$-body simulations including scalar fields and related models have been
performed before (Linder \& Jenkins 2003; Mainini~et~al.  2003;
Macci\`o~et.~al. 2004; Springel \& Farrar 2007; Kesden \& Kamionkowski
2006a, 2006b; Farrar \& Rosen 2007; Baldi~et~al.  2008; Oyaizu 2008;
Keselman~et~al. 2009; Li \& Zhao 2009). For example, in the work of
Macci\`o~et~al.~(2004) the simulations included several effects due to
the coupling between dark energy and dark matter (e.g. modified
gravitational constant, an extra dragging term in Newton's equations
and time variable dark matter particle masses), but did not consider a
spatial variation of the dark energy scalar field.  The more complete
simulation of the scalar field by Li \& Zhao (2009) shows that this
approximation is only good for a limited choice of parameters
and the scalar field potential. Here we extend the work of Li \& Zhao
(2009).

This paper is organized as follows: In \S~\ref{sect:description},
we shall briefly review the general equations of motion for the
coupled scalar field model introduced in Li \& Zhao (2009), and
present our specific choices of the coupling function and the scalar
field potential. In \S~\ref{sect:non-linear}, we 
describe the formulae and the algorithm of the
$N$-body simulation, analyze the results of our coupled scalar field $N$-body simulations, compare it
with that of the standard $\Lambda$CDM model, and explain the
physical origin of the new features. Finally, we conclude and discuss observational implications in
\S~\ref{sect:disc}.

\section{The Coupled Scalar Field Model}

\label{sect:description}

\subsection{The Model}

\label{subsect:Model}

All properties of our coupled scalar field model can be derived from minimizing the action 
associated with the following Lagrangian density (the index $a$ runs from $0$ to $3$):
\begin{eqnarray}\label{eq:Lagrangian}
\mathcal{L} &=&
\left[\frac{R}{2} - \frac{1}{2}\nabla^{a}\varphi\nabla_{a}\varphi + V_{\rm eff}(\varphi) \right],
\end{eqnarray}
which includes the Ricci scalar $R$, and a dimensionless scalar field $\varphi$ with a kinetic and an effective potential term.
The latter is given by
\begin{eqnarray}\label{eq:potential}
V_{\rm eff}(\varphi) & \equiv & V(\varphi)-  \kappa(\varphi) \mathcal{L}_{\mathrm{CDM} }
\end{eqnarray}
where the potential and the coupling function $\kappa(\varphi)$ are controlled by two dimensionless parameters, $\mu$ and $\gamma$, respectively.
More rigorously, the potential $V(\varphi)$ is
\begin{equation}
V(\varphi) = \Lambda_{0} {\left[1-\exp (-\varphi )\right]^{-\mu}}
\label{eq:phi_potential}
\end{equation}
and the coupling function $\kappa(\varphi) \equiv 8\pi G \exp (\gamma\varphi )$, as given in Li \& Zhao (2009),
where $\Lambda_{0}$ is a constant on the order of the cosmological constant, and $G$ is Newton's constant of gravitation. Considering the non-relativistic, weak field limit of Eq. (\ref{eq:potential}),
\begin{equation}
V_{\rm eff}(\varphi) \approx  \Lambda_0 \varphi^{-\mu} +  8\pi G (1+ \gamma \varphi ) \rho_{\rm CDM},
\end{equation}
the meaning of this particular parameterization can be understood as follows:
As the scalar field $\varphi$ tends to minimize the effective potential, the potential term $\Lambda_0 \varphi^{-\mu}$ and the coupling $(1+ \gamma \varphi)$ to the CDM density ($\mathcal{\rho}_{\mathrm{CDM}} \sim -\mathcal{L}_{\mathrm{CDM}}$ in the non-relativistic, weak-field limit) lead to competing effects, favoring smaller and larger values of $\varphi$, respectively. \footnote{The dark matter Lagrangian $\mathcal{L}_{\mathrm{CDM}}$ specifies
the geodesic flow for many point-like particles of four-velocity $u_a$ and density $\rho_{\rm CDM}$}
The balance of these two effects, minimizing the effective potential $V_{\rm eff}$, is controlled by the two dimensionless
parameters $\mu$ and $\gamma$: $\mu$ is very small and controls the time when the effect of the
scalar field (mainly exerting the finite-ranged fifth force on
dark matter particles on galaxy cluster scales) becomes important
for cosmology while $\gamma$ determines how large it will
ultimately be (Li \& Zhao 2009).
More specifically, the scalar field equation of motion is
\begin{eqnarray}\label{eq:phiEOM}
\square\varphi + \frac{\partial V(\varphi)}{\partial\varphi} +
\rho_{\mathrm{CDM}}
8\pi G \gamma \exp (\gamma \varphi ) &=& 0.
\end{eqnarray}

Einstein's equations can be expressed as
\begin{eqnarray}\label{eq:EMT_tot}
\frac{1}{8\pi G} G_{ab} &=&  \exp (\gamma\varphi ) \rho_{\mathrm{CDM}}u_{a}u_{b} + T^{\varphi}_{ab},
\end{eqnarray}
where $G_{ab}$ is the Einstein tensor, and  the right hand side is
the energy-momentum tensor of the the scalar field and CDM with a four-velocity $u^a$; the scalar field's  is given by
\begin{eqnarray}\label{eq:phiEMT}
8\pi G T^{\varphi}_{ab} &=& \nabla_{a}\varphi\nabla_{b}\varphi -
g_{ab}\left[\frac{1}{2}\nabla_{c}\varphi\nabla^{c}\varphi -
V(\varphi)\right].
\end{eqnarray}
Note that the energy-momentum tensors for the scalar field $\varphi$ and the dark matter
are not individually conserved due to their coupling, whereas their sum is.

Eqs.~(\ref{eq:phiEOM}, \ref{eq:EMT_tot}) summarize all the physics that will be used in our analysis. An immediate application is
the prediction of a uniform Hubble expansion. The model's expansion is completely indistinguishable from $\Lambda$CDM for values of $\gamma \sim \mathcal{O}(1)$ and $\mu \ll 1$; the actual difference is on the order of $\mathcal{O}(\mu)$. Basically, this is due to the large enough scalar's mass, forcing the field near the potential minimum, which itself is almost time-independent for $\mu\ll 1$. A quantitative explanation is
given in Li \& Zhao (2009). We now proceed to break the degeneracy via nonlinear clustering.

\begin{figure}[t]
\centerline{
\epsfysize=2.0in \epsffile{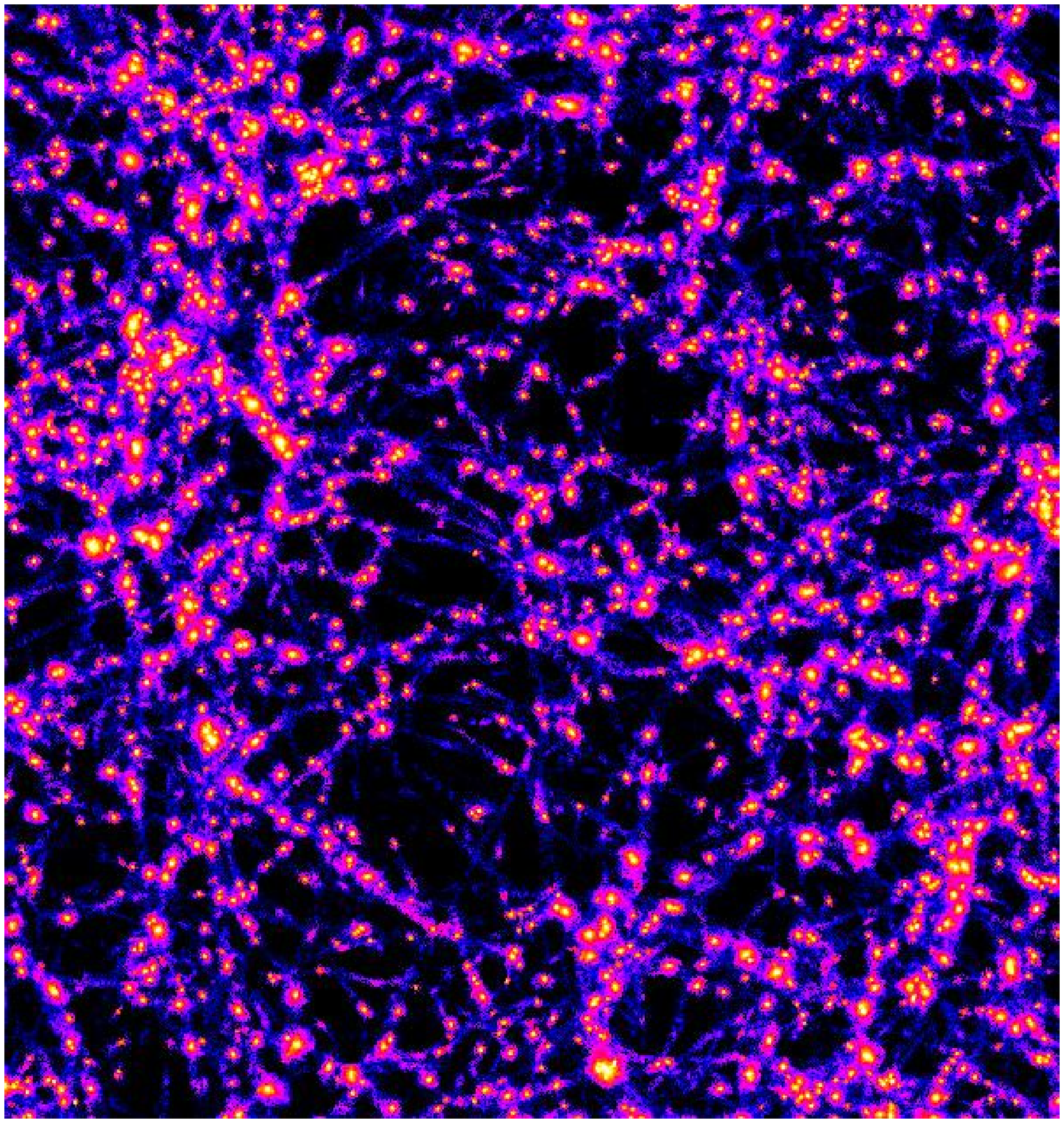}
\epsfysize=2.0in \epsffile{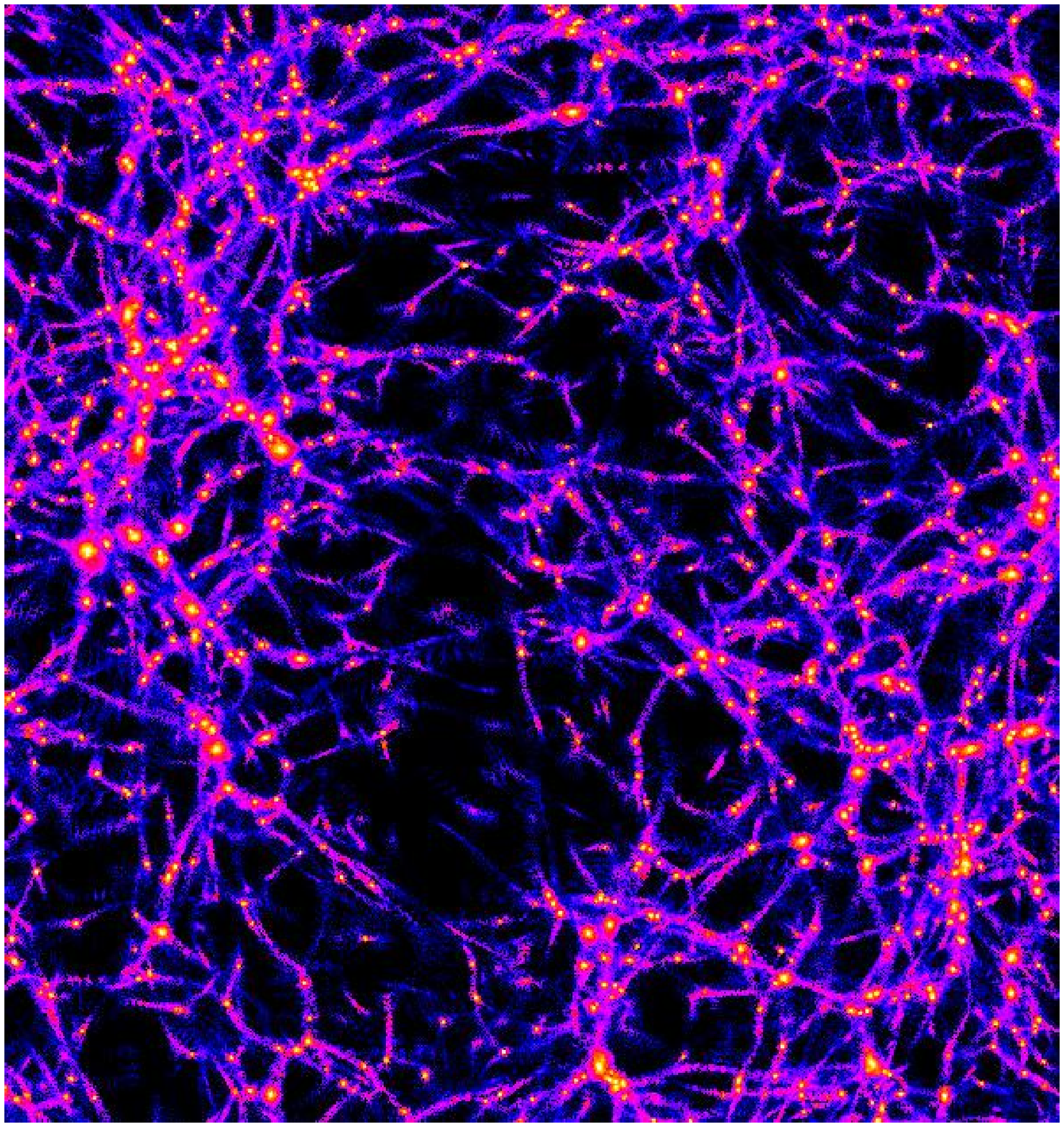}
}
\caption{\scriptsize Overdensity fields at $z=0$ for the $\varphi$CDM model with $\gamma=1$, $\mu=10^{-5}$ (left) and the $\Lambda$CDM model (right).  The former has developed more small-scale structure within the void.  }\label{fig:density}

\end{figure}


\subsection{The Nonrelativistic Equations}

\label{subsect:noneqn}

The first step towards a numerical simulation is to simplify the
relevant equations of motion in the non-relativistic and quasi-static
limit (in the sense that the time derivatives can be safely
neglected compared with the spatial derivatives).

Li \& Zhao (2009) showed that the scalar equation of motion,
Eq.~(\ref{eq:phiEOM}) and the Poisson equation can be simplified as
\begin{eqnarray}\label{eq:WFphiEOM}
\frac{ \partial_{\mathbf{x}}^{2}\varphi}{a^2} &\approx&
8\pi G \gamma  \left[\rho_{\mathrm{CDM}} -
\bar{\rho}_{\mathrm{CDM}}  \right]
-\mu \Lambda_0 \left[\varphi^{-\mu-1}-\bar{\varphi}^{-\mu-1}\right] \\
\label{eq:WFPoisson}
\frac{\partial_{\mathbf{x}}^{2}\Phi}{a^{3}} &\approx& 4\pi G
\left[\rho_{\mathrm{CDM}}
-\bar{\rho}_{\mathrm{CDM}}  \right] - 
\Lambda_0 \left[\varphi^{-\mu}-\bar{\varphi}^{-\mu}\right].
\end{eqnarray}
Note that the above two equations have similar source terms, partly from matter and partly from the
scalar field.\footnote{
The notation $\partial^{2}_{\mathbf{x}}=-\vec{\nabla}_{\mathbf{x}}^{2} =
\partial^{2}_{x}+\partial^{2}_{y}+\partial^{2}_{z}$
is defined with respect to the comoving coordinate $\mathbf{x}$
such that $\vec{\nabla}_{\mathbf{x}}=a\vec{\nabla}_{\mathbf{r}}$,
where $a$ is the usual scale factor of the Universe. In the following, $\bar{\varphi}$ and
$\bar{\rho}_{\mathrm{CDM}}$ denote the background values of $\varphi$ and $\rho_{\mathrm{CDM}}$, respectively.
Although we have used the approximation $\rho_{\rm CDM} \exp (\gamma \varphi) \sim \rho_{\rm CDM}$ for a simpler presentation, we keep the factor
$\exp (\gamma \varphi)$ as well as the potential given by Eq.~(\ref{eq:phi_potential}) in the actual simulation.}

Finally, the equations of motion of the dark matter
particles are also modified as
\begin{eqnarray}\label{eq:WFdxdtcomov}
\frac{d\mathbf{x}}{dt} &=& \frac{\mathbf{p}}{a^{2}},\\
\label{eq:WFdpdtcomov} \frac{d\mathbf{p}}{dt} &=&
-\frac{1}{a}\vec{\nabla}_{\mathbf{x}}\Phi -
\gamma  \vec{\nabla}_{\mathbf{x}}\varphi ,
\end{eqnarray}
where the canonical momentum conjugate to the comoving
coordinates $\mathbf{x}$ is $\mathbf{p}=a^{2}\dot{\mathbf{x}}$.
Note that the two terms on the right hand side of
Eq.~(\ref{eq:WFdpdtcomov}) correspond to gravity and fifth force, respectively (Li \& Zhao 2009).  The scalar field $\varphi$
is on the order of magnitude of $\mu$, comparable to the dimensionless potential $\Phi$.
Eqs.~(\ref{eq:WFphiEOM}, \ref{eq:WFPoisson}, \ref{eq:WFdxdtcomov},
\ref{eq:WFdpdtcomov}) are used in the code to evaluate the forces on
the dark matter particles and to evolve their positions and momenta in time.

The validity and limitation of the approximation present in the above equations, in particular neglecting the time derivatives, have been extensively discussed in Li \& Zhao (2009). We emphasize that these approximations do not hold in linear regime where
the scalar field's time dependence is essential for structure growth.  However such terms have indeed been shown to be negligible
on scales much smaller than the horizon scale (Li \& Zhao 2009; Oyaizu 2008). To make our predictions more quantitative and rigorous
compared to previous analyses (Macci\`o~et.~al. 2004; Kesden \& Kamionkowski 2006a, 2006b; Farrar \& Rosen 2007), we now analyze the first $N$-body simulations in the above framework. Considering the linear regime, Li \& Zhao (2009) have already been able to constrain the
parameters $\mu$ and $\gamma$ to a fairly narrow range. Here we set $\gamma$ on the order of unity to force a significant ratio of the fifth
force to gravity ($\sim 2\gamma$), and explore the range $10^{-7} \leq\mu\leq 10^{-5}$, covering three orders of magnitude.
Restricting ourselves to the above should suffice as the model is either essentially indistinguishable from $\Lambda$CDM or deviates too
much from it (already at the linear level) beyond this parameter space, thus being of no further interest (Li \& Zhao 2009).

\section{Nonlinear Structure Formation}

\label{sect:non-linear}

In this section, we present some results of the first $N$-body runs and
describe the qualitative behaviour of the coupled scalar field model.

\subsection{The N-Body Code}

\label{subsect:N-body}

We adapt the Multi-Level Adaptive Particle Mesh (MLAPM) code (Knebe et
al.~2001) to include the scalar field, and its coupling to the dark
matter $N$-body particles. One benefit of the adaptive scheme is that
the majority of computing resources is dedicated to few high
density regions to ensure higher resolution, which is desirable since
we expect the behaviour of the scalar field to be more complex there.

The main modifications to the MLAPM code for our model are:
\begin{enumerate}
    \item We have added a parallel solver for the scalar field based
      on Eq.~(\ref{eq:WFphiEOM}). The solver uses a similar nonlinear
      Gauss-Seidel method (Briggs et al.~2000, Press et al.~1992) and
      the same criterion for convergence as the Poisson solver.
    \item The resulting value for $\varphi$ of the first step is
      used to calculate the local mass density of the scalar field and
      thus the source term for Poisson's equation, which is solved
      using a fast Fourier transform to obtain the local gravitational
      potential $\Phi$ [cf.~Eq.~(\ref{eq:WFPoisson})].
    \item The fifth force is obtained by differentiating $\varphi$,
      and the gravitational force is calculated by differentiating $\Phi$,
      as in Eqs.~(\ref{eq:WFdxdtcomov}, \ref{eq:WFdpdtcomov}).
    \item The momenta and positions of particles are then updated,
    taking into account both gravity and the fifth force, just as in normal $N$-body codes.
\end{enumerate}
More technical details on the code, as well as how
Eqs.~(\ref{eq:WFphiEOM}, \ref{eq:WFPoisson}, \ref{eq:WFdxdtcomov},
\ref{eq:WFdpdtcomov}) are incorporated into MLAPM using its own
internal units, have been given in Li \& Zhao (2009) and will not
be presented here.

\subsection{Numerical Results from the $N$-body Runs}

\label{subsect:numresult}

We have performed $6$ runs of the modified code with parameters
$\gamma=0.5, 1$ and $\mu=10^{-5}, 10^{-6}, 10^{-7}$, respectively. For
all these runs, there are $128^{3}$ dark matter particles, and the
simulation box size is chosen as $B = 64h^{-1}~\mathrm{Mpc}$, with
$h$ being the usual dimensionless Hubble parameter and $128$ domain grid cells in each
direction. We assume a $\Lambda$CDM background cosmology which is a
very good approximation for $\mu\ll1$ (Li \& Zhao 2009); in addition, we adopt
present values for the fractional energy densities of dark matter and dark energy,
$\Omega_{\mathrm{CDM}}=0.28$ and $\Omega_{\Lambda}=0.72$, and the
normalization of the power spectrum is chosen as $\sigma_8=0.88$. Note that the simulation does only take dark matter into
account, baryons will be added in a forthcoming work to study
the bias effect caused by the dark matter coupling. Given these
parameters, the mass and spatial resolution of the simulation are
$9.71\times10^{9}~M_{\bigodot}$ and $\sim23.44h^{-1}~\mathrm{kpc}$
(for the most refined regions), respectively.  This spatial resolution
in high density regions is necessary and sufficent to precisely probe
the scalar field in regions where the fifth force is considerably short-ranged.

All simulations started at redshift $z=49$. In principle, modified
initial conditions, i.e. the initial displacements and velocities of particles
which are obtained from a given linear matter power spectrum, need to be
generated for the coupled scalar field model because the Zel'dovich
approximation (Efstathiou et al.~1985) is also affected by the scalar
field coupling. In practice, however, we find that the effect on the
linear matter power spectrum at this high redshift is negligible
($\lesssim\mathcal{O}(10^{-4})$) for our choice of the parameters
$\gamma$ and $\mu$. Thus we simply use the $\Lambda$CDM initial
displacements/velocities for the CDM particles in our simulations,
which are generated using GRAFIC (Bertschinger 1995), again using
$\Omega_{\mathrm{CDM}}=0.28$, $\Omega_{\Lambda}=0.72$ and
$\sigma_{8}=0.88$.  An example of the final density field at redshift $z=0$ is
shown in Fig.~\ref{fig:density} for comparison with the $\Lambda$CDM
simulation.

We look for all virialized isolated haloes within our
computational volume using a Spherical Overdensity algorithm. For this purpose, we
employ a time varying virial density contrast which is determined using the
fitting formula presented in Mainini \etal (2003), and adopt the
same virial density contrast for all models. In addition, we include all haloes with more than $200$ particles
into the halo catalogue (see Macci\`o \etal 2008 for further details on our halo
finding algorithm). Power spectra have been computed through a
(fast) Fourier transform of the matter density field, computed on
a regular grid $N_G\times N_G\times N_G$ from the particle
distribution via a Cloud-in-Cell algorithm (see Casarini \etal 2009).
We set $N_G=256$ which gives a maximum mode of $k\approx 20h$Mpc$^{-1}$ well above the simulation
resolution.

\begin{figure}[t]
\centerline{\epsfxsize=3.3in \epsffile{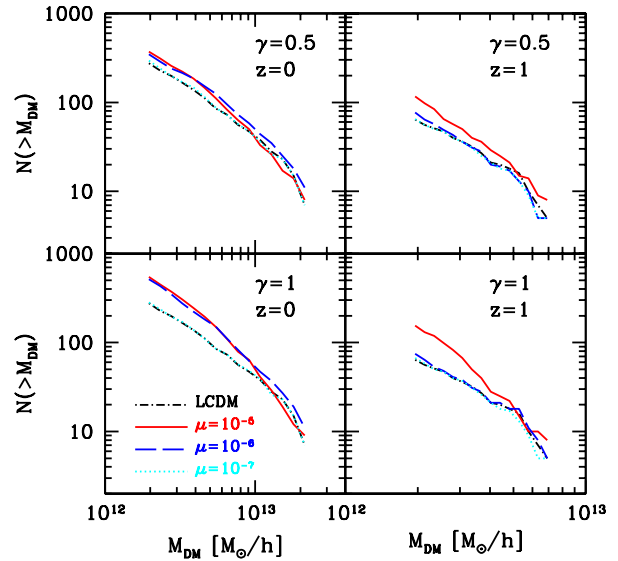}}
\caption{\scriptsize Mass functions for $\gamma=0.5$ (upper panels)
and  $\gamma=1$ (lower panels) for different values of $\mu$
at $z=0$ and $z=1$. The $\Lambda$CDM mass function is also plotted as a
(black) dot-dashed curve for comparison.} \label{fig:MFall}
\end{figure}

\begin{figure}[t]
\centerline{\epsfxsize=3.2in \epsffile{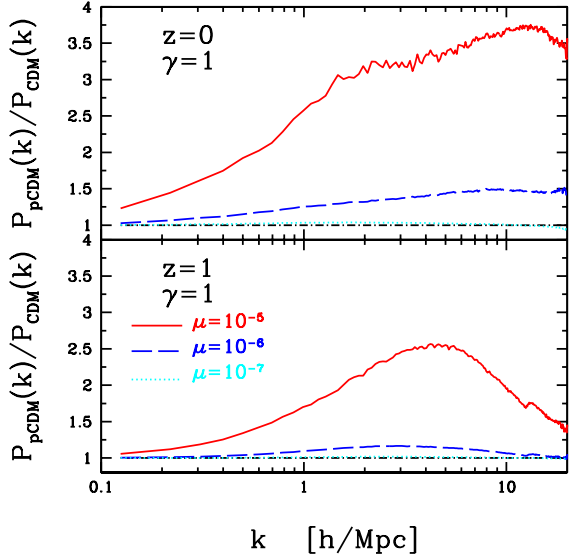}}
\caption{\scriptsize Ratios of calculated nonlinear matter power
spectra for $\gamma=1$ and $\mu=10^{-5}$ (red),
$10^{-6}$ (blue) and $10^{-7}$ (green) as well as for
that of $\Lambda$CDM. Shown are results for two redshifts, $z=1, 0$. 
At large scales (small $k$) the curves converge to the horizontal curves (identical to $1$, black dotted).
Note that, using analytic results, the difference is expected to be small 
on both large and very small scales, and decreases at higher redshift.  Error bars of future lensing observations
are likely small enough to detect any deviation from $\Lambda$CDM on intermediate scales ($k=0.1-10h$Mpc$^{-1}$) at a $30$\% level.}
\label{fig:PSgamma1}
\end{figure}

\begin{figure}[t]
\centerline{\epsfxsize=3.2in \epsffile{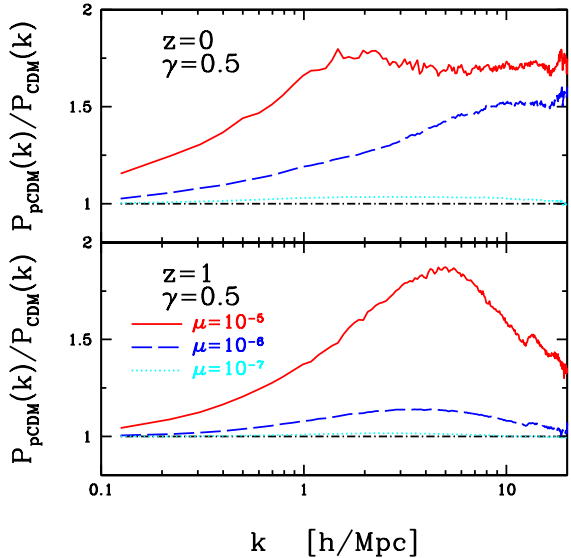}}
\caption{\scriptsize The same as Fig.~\ref{fig:PSgamma1}, but for
$\gamma=0.5$.} \label{fig:PSgamma0.5}
\end{figure}

In Fig. ~\ref{fig:MFall} we show
the mass functions for the runs with $\gamma=1.0, 0.5$ and
$\mu=10^{-5}, 10^{-6}, 10^{-7}$ and the fiducial $\Lambda$CDM
simulation at two output redshifts $z=1$ and $0$. The
nonlinear matter power spectra of these models are displayed
in Figs.~\ref{fig:PSgamma1} and \ref{fig:PSgamma0.5}, respectively.

\subsection{Interpretation of the Results}

The results of the $N-$body simulations can be understood
intuitively, as we shall discuss below. In general, a scalar field coupled to
matter particles produces a fifth force
[cf.~Eq.~(\ref{eq:WFdpdtcomov})] on the latter, which has a finite
range $m_\varphi^{-1}$ determined by the mass $m_{\varphi}$ of the scalar field. If
$m_{\varphi}$ is small and almost constant across space then the fifth-force
effect essentially leads to an increase in the effective gravitational
constant which governs structure formation (Macci\`o et
al. 2004).  Li \& Zhao (2009) have shown that for certain regions of parameter space
and specific choices of the potential, this is indeed a good
approximation. Mathematically this corresponds to neglecting the source terms starting with 
$\Lambda_0$  in Eqs. (\ref{eq:WFphiEOM},\ref{eq:WFPoisson}), hence 
the fifth force $\gamma \nabla \varphi$ is about a factor $2 \gamma^2$ times the gravitational force $\mathbf{\nabla}\Phi/ a$.

In another situation, when the scalar field has a very steep potential,
$m_{\varphi}$ depends sensitively on the local matter density
(Khoury \& Weltman 2004) so that it almost resides at the minimum of
its effective potential
\begin{equation}
V_{eff}(\varphi)=V(\varphi)+ 8\pi G\rho_{\mathrm{CDM}}\exp (\gamma\varphi)
\end{equation}
throughout space, i.e., $\varphi \sim \Lambda_0 \mu / (8 \pi G \rho_{\rm CDM})$.  
This is known as the chameleon effect whose direct consequence
is that in a high density environment, $m^{2}_{\varphi}$ gets very heavy,
\begin{equation}
m^{2}_{\varphi}=\partial^{2}V_{eff}/\partial\varphi^{2} = \Lambda_0 \mu (1+\mu) \varphi^{-\mu-2} \propto \mu^{-1} \rho^{-2}_{\rm CDM},
\end{equation}
and the fifth force becomes very short-ranged, with its effect being suppressed due to $\gamma \nabla \phi \propto \gamma \mu \nabla \rho_{\rm CDM}^{-1}$. In general the
smaller $\mu$ is and/or the larger $\gamma, \rho_{\mathrm{CDM}}$
are, the heavier becomes $m_{\varphi}$ and thus the stronger the
chameleon effect will be. Furthermore, since the value of
$\varphi$ inside a region also depends on its boundary condition,
which in our case matches the background $\bar{\varphi}$
asymptotically, we see that a smaller $\bar{\varphi}$ leads to
a smaller $\varphi$ and a heavier $m_{\varphi}$, and therefore to a stronger
chameleon effect.

There are several interesting features in Fig.~\ref{fig:MFall} which
can be understood schematically. First of all, our models produce more
halos within the considered mass range than $\Lambda$CDM due to the enhancing effect of the fifth
force. Secondly, a smaller $\mu$ means that the fifth force is more
severely suppressed by the chameleon effect, and thus causes a small
deviation from $\Lambda$CDM. Thirdly, a larger $\rho_{\mathrm{CDM}}$
also means that the fifth force is more severely suppressed, and this
is why at high redshifts the deviation from $\Lambda$CDM (for the same
$\gamma$ and $\mu$) is smaller. Fourthly, the influence of the parameter $\gamma$ is more
complicated: A larger $\gamma$ will strengthen the chameleon effect,
tending to suppress the fifth force, but at the same time it increases the magnitude of the fifth force. In cases where
chameleon effect is weak (e.g., $\mu=10^{-5}$), however, we do see that a larger
$\gamma$ leads to larger deviations from $\Lambda$CDM.

Also note that the deviation from the $\Lambda$CDM mass function is
more significant towards the low-mass end. To understand why this is the case,
consider a mass range $[M_0, M_0+\Delta M]$. At a certain redshift,
some halos which should have been in this range in $\Lambda$CDM indeed fall into the mass range $(M_0+\Delta M,
\infty)$ in our model as the fifth force accelerates the formation of structures
(this tends to reduce the number of halos with mass $>
M_0$ as two halos which are separated in $\Lambda$CDM merge into
one here), while some halos which should have been in the mass
range $<M_0$ in $\Lambda$CDM actually fall in the mass range
$[M_0, M_0+\Delta M]$ in our model (this increases the number of
halos with mass $> M_0$). This effect is weaker for the largest halos
because of competing effects due to merging of small halos.
As the mass increases, the difference between the mass functions of the
two models narrows down.

In the matter power spectrum, we see something similar: Smaller $\mu$
and larger $\rho_{\mathrm{CDM}}$ (higher redshift) severely suppress the fifth
force and lead to smaller deviations from
$\Lambda$CDM; increasing $\gamma$ strengthens the fifth force,
thereby causing large deviations from $\Lambda$CDM. Interestingly, the
deviation becomes largest on intermediate scales: large scales are
beyond the probe of the fifth force, and thus not significantly affected,
while the density on small scales is high and the fifth force is
suppressed.

\section{Conclusion}

\label{sect:disc}

We have presented a general framework to study nonlinear structure formation in coupled scalar
field models, in particular the models of Li \& Zhao (2009).
While these models are virtually indistinguishable from $\Lambda$CDM on both very large and very
small scales, intermediate scales at low redshift ($z\lesssim 1$) relevant for galaxy clusters ($\sim 10^{2}-10^{3}$kpc)
open a new window to test and constrain the interesting part of the parameter space.

On these scales, the matter power spectrum is significantly increased compared to that of $\Lambda$CDM.
Observationally, this would most likely appear as a change of $\sigma_{8}$ on the order of $15$-$20$\% for models
with $\gamma=0.5-1$ and $\mu=10^{-6}$ (see Fig.~2).
Any variation of $\sigma_8$ seems to be lower than 30\% for current lensing measurements such as
the CFHT Legacy Survey (e.g., Hoekstra et al. 2006; Fig.~11 of Fu et al. 2008) over a rather limited range;
however, future surveys, such as the Kilo-Degree Survey (KIDS), will be able to measure the scale dependence within the range
$k=0.1-10h$Mpc$^{-1}$, where the deviation of the models from $\Lambda$CDM is maximal.




\section*{Acknowledgments}

This work has been performed within the HPC-EUROPA project, with the
support of the European Community Research Infrastructure Action
under the FP8 "Structuring the European Research Area" Programme.
We thank Simon-Porteges Zwart and Marco Baldi for discussions.  
H. Zhao thanks for financial support from the Dutch NWO visitorship No.040.11.089 to Henk Hoekstra, B.~Li acknowledges financial support from an UK Overseas Research Studentship, the Cambridge Overseas Trust and Queens' College Cambridge. M.~Feix is supported by a scholarship from the Scottish Universities Physics Alliance (SUPA).

\end{document}